\documentclass{webofc}

\usepackage[varg]{txfonts}   
\usepackage{hyperref}
\usepackage{units}
\usepackage{xcolor}
\usepackage{ulem} 

\usepackage{url}
\hypersetup{colorlinks=true,citecolor=blue,urlcolor=blue,linkcolor=blue}
\begin{document}
\begin{flushright}
\texttt{DESY-26-038}\\
\texttt{IFT–UAM/CSIC-26-29}\\
\texttt{FR-PHENO-2026-006}
\end{flushright}
\vspace{-2cm}\includegraphics[height=3cm]{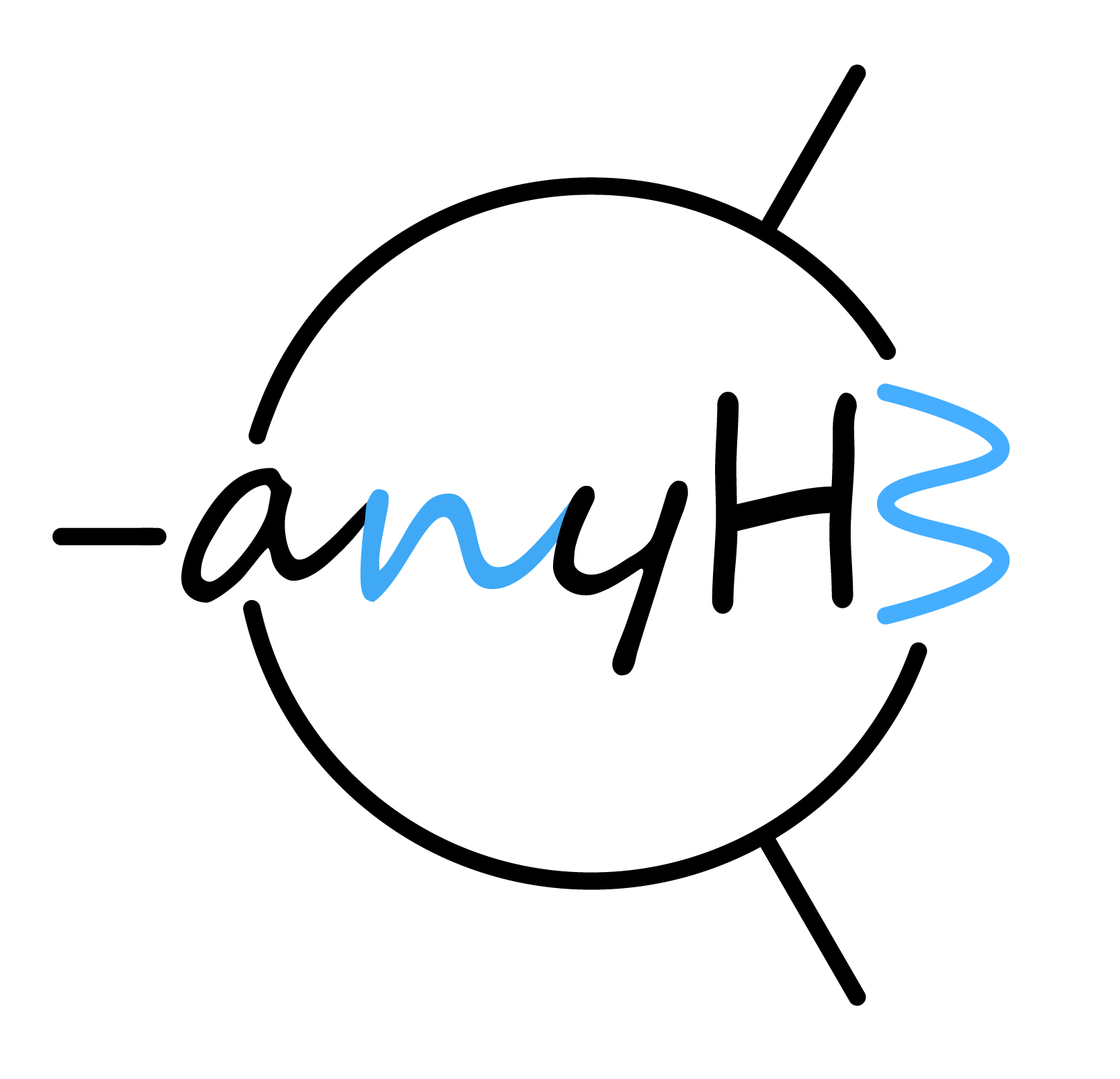}

\title{\vspace{-1cm}Precise predictions for trilinear Higgs couplings and Higgs pair production in extended scalar sectors with \texttt{anyH3} and \texttt{anyHH}}
%
%

\author{\firstname{Henning} \lastname{Bahl}\inst{1}\fnsep\thanks{\email{bahl@thphys.uni-heidelberg.de}} \and
        \firstname{Johannes} \lastname{Braathen}\inst{2}\fnsep\thanks{\email{johannes.braathen@desy.de}} \and
        \firstname{Martin} \lastname{Gabelmann}\inst{3}\fnsep\thanks{\email{martin.gabelmann@physik.uni-freiburg.de}} \and
        \firstname{Kateryna} \lastname{Radchenko}\inst{4}\fnsep\thanks{\email{kateryna.radchenko@desy.de}} \and 
        \firstname{Georg} \lastname{Weiglein}\inst{2,5}\fnsep\thanks{\email{georg.weiglein@desy.de}}
        }

\institute{
Institut für Theoretische Physik,Universit{\"a}t Heidelberg, Philosophenweg 16, 69120, Heidelberg, Germany
\and
Deutsches Elektronen-Synchrotron DESY, Notkestr.~85,  22607  Hamburg,  Germany
\and 
Albert-Ludwigs-Universit{\"a}t Freiburg, Physikalisches Institut, Hermann-Herder-Str.~3, 79104, Freiburg, Germmany
\and
Instituto de F\'isica Te\'orica (UAM/CSIC), 
Universidad Aut\'onoma de Madrid, Cantoblanco, 28049, Madrid, Spain
\and
Institut für Theoretische Physik, Universität Hamburg, Luruper Chaussee 149, 22761, Hamburg, Germany
}

\abstract{
A central objective of future collider experiments is to probe the structure of the Higgs potential, which requires access to trilinear scalar couplings, in particular the self-coupling of the observed Higgs boson. While this coupling is fixed in the Standard Model (SM), it can receive sizable modifications in many Beyond the SM (BSM) scenarios, often connected to solutions of open problems such as the origin of the matter–antimatter asymmetry of the Universe. 
In theories with extended scalar sectors, radiative corrections involving additional scalar states can significantly affect both the Higgs self-coupling and other trilinear scalar interactions, with important consequences for predictions of physical observables. Precise theoretical calculations are therefore essential for the interpretation of precision Higgs measurements and for identifying indirect signatures of new physics. This contribution presents the latest version of the public tool \texttt{anyBSM}, which provides automated calculations of all trilinear scalar couplings at full one-loop order in arbitrary renormalisable theories, including full momentum dependence and flexible renormalisation-scheme choices. In addition, the new module \texttt{anyHH} for di-Higgs production in gluon fusion is discussed in several exemplary BSM models, including scenarios with multiple resonances.
}
\maketitle
\pagestyle{plain}

\noindent\textit{Talk presented at the International Workshop on Future Linear Colliders 2025 (LCWS2025), Valencia, Spain}
\newpage

\section{Introduction}
\label{intro}
\setcounter{page}{1}

Di-Higgs production can serve as a crucial probe of the Higgs potential, as it provides direct access at leading order to the trilinear Higgs self-coupling of the 125-GeV Higgs boson discovered at the Large Hadron Collider (LHC)~\cite{Plehn:1996wb,ATLAS:2019mfr}. Through this measurement, it offers a key window into the mechanism of electroweak symmetry breaking and the nature of the electroweak phase transition. In addition to probing the Standard Model (SM) Higgs potential through the measurement of the Higgs trilinear self-interaction, di-Higgs production can be sensitive to resonant contributions from additional neutral scalar states predicted in physics Beyond the Standard Model (BSM), and thus provide information on BSM trilinear scalar couplings.

Current experimental searches target either non-resonant production, constraining deviations in the trilinear Higgs self-coupling $\lambda_{hhh}$, or resonant production of new scalar states. With Run 2 data, the ATLAS and CMS collaborations have set bounds on the normalized trilinear coupling modifier $\kappa_{\lambda}\equiv\lambda_{hhh}/\lambda_{hhh}^{\text{SM},\ (0)}$~\cite{CMS:2026nuu} (where $\lambda_{hhh}^{\text{SM},\ (0)}$ is the lowest-order SM prediction for the trilinear coupling), and future data from Run 3 and the High-Luminosity LHC (HL-LHC) are expected to significantly improve the sensitivity to di-Higgs production~\cite{ATLAS:2025eii}. These experimental efforts require equally precise theoretical predictions, particularly in view of potentially sizeable BSM effects.

While substantial progress has been made in reducing SM uncertainties, BSM loop effects\footnote{See also Refs.~\cite{Heinrich:2022idm,Heinrich:2023rsd,Heinrich:2024rtg} and~\cite{Buchalla:2018yce,Capozi:2019xsi,Heinrich:2020ckp,Braun:2025hvr,Brivio:2025sib} for works in the context of SMEFT and HEFT, respectively.} in trilinear Higgs couplings~\cite{Kanemura:2002vm,Kanemura:2004mg,Aoki:2012jj,Kanemura:2015fra,Kanemura:2015mxa,Hashino:2015nxa,Kanemura:2016lkz,Kanemura:2017wtm,Kanemura:2017gbi,Chiang:2018xpl,Basler:2018cwe,Senaha:2018xek,Braathen:2019pxr,Kanemura:2019slf,Braathen:2019zoh,Basler:2020nrq,Braathen:2020vwo,Bahl:2022jnx,Bahl:2022gqg,Bahl:2023eau,Aiko:2023nqj,Basler:2024aaf,Bahl:2025wzj,Braathen:2025qxf,Braathen:2025svl} and their impact on di-Higgs production~\cite{Bahl:2022jnx,Bahl:2023eau,Heinemeyer:2024hxa, Arco:2025pgx, Braathen:2025qxf} have only recently received increased attention.
Building on the public code \texttt{anyH3}~\cite{Bahl:2023eau}, which provides automated one-loop predictions for the trilinear Higgs self-coupling, we present several extensions of the overarching \texttt{anyBSM} framework. In particular, we generalise the calculations in \texttt{anyH3} to trilinear couplings with non-identical external scalar particles, thereby enabling precise predictions for interactions between all scalar states in a BSM framework. In addition, we introduce the new subpackage  \texttt{anyHH}, which calculates total and differential di-Higgs cross-sections in gluon fusion at the (HL-)LHC, consistently incorporating loop corrections to all relevant trilinear scalar couplings as well as the full momentum dependence therein.

\section{General trilinear scalar couplings at the one-loop level}

Models with extended scalar sectors generically feature a rich structure of trilinear self-interactions among multiple Higgs and Higgs-like states. While in the SM only the single coupling $\lambda_{hhh}$ exists, BSM theories typically predict a full set of interactions $\lambda_{h_i h_j h_k} \equiv \lambda_{ijk}$ among several neutral and/or charged scalar mass eigenstates. The extended version of \texttt{anyH3} automatically identifies all trilinear scalar interactions and computes the corresponding one-loop–corrected couplings in a fully general setup for a given \texttt{UFO} model file. 

The couplings are evaluated as momentum-dependent three-point functions including the full set of one-loop contributions: genuine vertex diagrams, tadpole terms (depending on the treatment of tadpoles chosen by the user), external-leg corrections, and the corresponding counterterms. The code is designed to allow flexible choices of renormalisation schemes. In particular, both $\overline{\text{MS}}$ and on-shell renormalisation for the masses and wave functions of all scalar fields are implemented in a fully automated way. Parameters specific to BSM models, such as vacuum expectation values or mixing angles, are treated in the $\overline{\text{MS}}$ scheme by default, but alternative counterterm definitions can be specified by the user independently for each parameter, and the corresponding vertex counterterms are generated automatically. The results have been validated through cross-checks with independent tools such as \texttt{FeynArts}~\cite{Kublbeck:1990xc,Eck:1992ms,Hahn:2000kx} and 
\texttt{FeynCalc}~\cite{Mertig:1991an,Shtabovenko:2016sxi,Shtabovenko:2020gxv} as well as with \texttt{BSMPT}~\cite{Basler:2018cwe,Basler:2020nrq,Basler:2024aaf} in the zero momentum approximation (for the latter).

\section{\texttt{anyHH}: Higgs pair production for scalar extensions of the SM}

We present the package \texttt{anyHH} for the calculation of the Higgs pair production cross section in gluon fusion ($gg \to hh$) at the leading order in QCD for any renormalisable extensions of the SM (with the only restriction of all BSM states being $\text{SU}(3)_c$ singlets). Additionally, the code incorporates one-loop corrections to the trilinear scalar couplings involved in the process, including full momentum dependence. This is of particular relevance given that these constitute the dominant BSM correction to the process in scenarios with large scalar couplings~\cite{Bahl:2022jnx}. These corrections can be 
included by means of an effective $\lambda_{ijk}$ coupling, by default in the zero momentum approximation. Optionally the full one-loop-corrected triangle form factors including momenta can be incorporated. Additionally, loop corrections to the $s$-channel propagator are included by default but can be switched off. Leading order decay widths are computed in \texttt{anyHH} using the optical theorem, or alternatively, can be provided by the user.

So far we only include the top-quark loop for the production, which is the dominant diagram in most of the scenarios. The code allows for an evaluation of both the differential di-Higgs cross section as a function of the partonic centre-of-mass energy $\sqrt{\hat{s}}$ as well as the total  di-Higgs cross section integrated over $\hat{s}$ convoluted with the proton PDFs using LHAPDF~\cite{Buckley:2014ana}. The resonant and non-resonant parts of the process can be computed separately. Our results were extensively cross checked against existing results in the literature, in particular against the SM  and a Two-Higgs-Doublet Model (THDM) using the public tool \texttt{HPAIR}~\cite{Plehn:1996wb,Dawson:1998py,Grober:2017gut,Abouabid:2021yvw} and against a singlet extension of the SM~\cite{Dawson:2015haa}. Finally, the code is not restricted to the $hh$ final state but it also straightforwardly predicts the total and differential cross-sections for the production of any pair of neutral scalars
$h_ih_j$, which may differ from each other,
via gluon fusion.\footnote{We note however that for the moment we are not including vector-boson contributions in the s-channel (i.e.\ Drell-Yan--like di-Higgs production) in the case of CP-odd final states, which should be compared to the gluon-fusion production depending on the scenario.}

\section{Phenomenological applications and new results}

As a first example of a phenomenological application we study the effects of loop induced couplings in the THDM for the scenario of the alignment limit~\cite{Gunion:2002zf} at tree level. In this limit, the coupling between three SM-like Higgs bosons, $\lambda_{hhh}$, is identical to its SM prediction at tree-level, and the coupling involving two SM-like Higgs boson and one heavier BSM scalar, $\lambda_{hhH}$, vanishes at lowest order. These are the only two trilinear scalar couplings involved in the pair production of $hh$ in the THDM. Since the tree-level condition of alignment is not protected by symmetry, it can be lifted at one-loop level, thus generating a non-zero value of $\lambda_{hhH}$ and in turn a resonant contribution to di-Higgs production.

\begin{figure}
    \centering
    \includegraphics[width=0.49\textwidth]{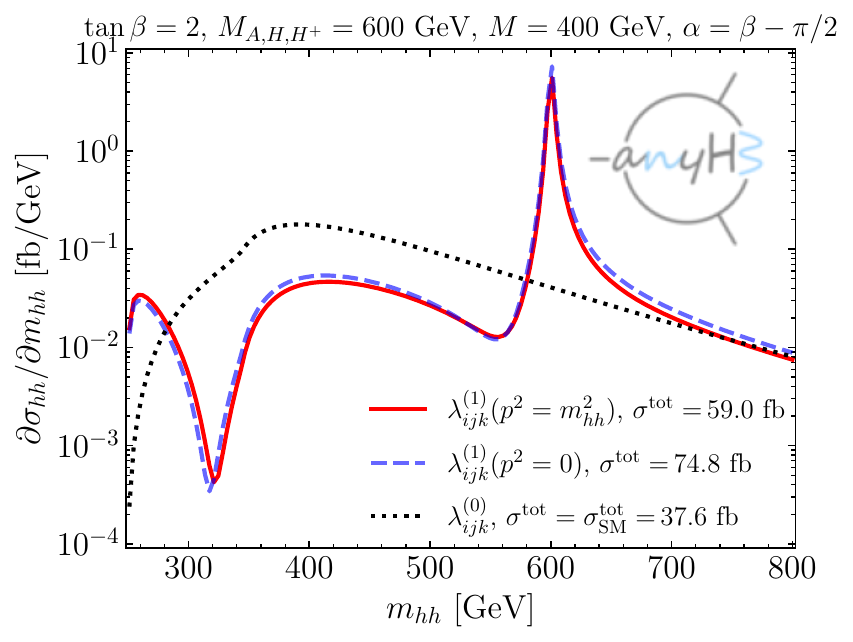}
    \includegraphics[width=0.49\textwidth]{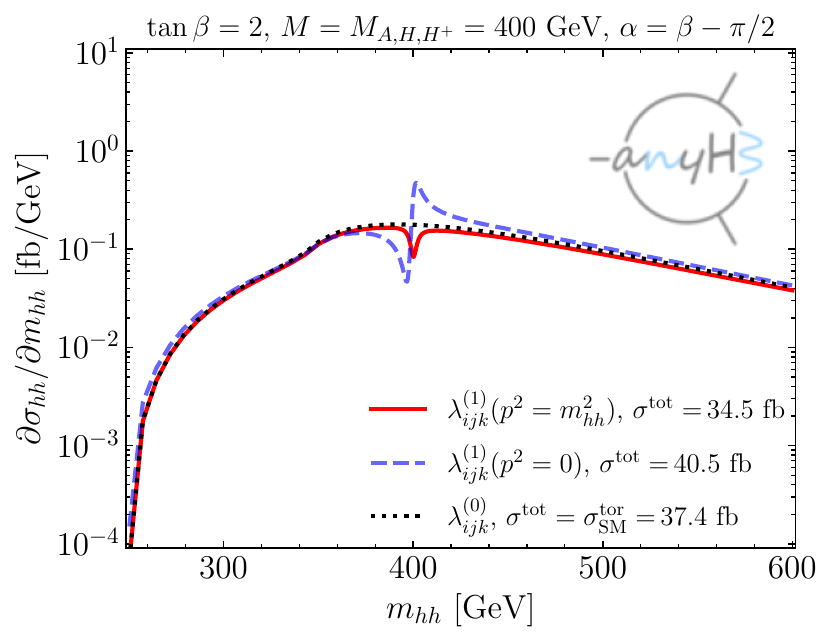}
    \caption{Differential cross-section for Higgs pair production in the tree-level aligned THDM type-II. \textbf{Left:} Comparison of the differential cross-section calculated using tree-level trilinear couplings (black dotted) and loop-corrected trilinears with (red solid) as well as without momentum dependence (blue dashed). \textbf{Right:} Same as upper left, but all mass parameters set to 400 GeV. The values of the trilinear scalar couplings for the scenario with $M_{A,H,H^\pm}=600\text{ GeV}$ and $M=400\text{ GeV}$ are $\lambda_{hhh}^{(1)}=$ 335.09 GeV and $\lambda_{hhH}^{(1)}=$ 191.78 GeV. 
    For the scenario with $M_{A,H,H^\pm}=M=400\text{ GeV}$ they are $\lambda_{hhh}^{(1)}=$ 176.29 GeV and $\lambda_{hhH}^{(1)}=$ 13.20 GeV. 
    }
    \label{fig:THDM_HHxs_differential}
\end{figure}

In Fig.~\ref{fig:THDM_HHxs_differential} we investigate these effects in specific scenarios of the THDM (defined in the title above each subplot), for details of the model and the free parameters see Refs.~\cite{Gunion:1989we,Lee:1973iz,Aoki:2009ha,Branco:2011iw}. We renormalise the model using the KOSY scheme~\cite{Kanemura:2004mg}. In the following we show the leading order result in QCD with a $K$-factor of 2 applied in order to approximately account for higher order QCD effects, which should be a reasonable assumption, especially given the large theoretical uncertainties in the predictions, \textit{c.f.} Ref.~\cite{Baglio:2018lrj}. In the upper left panel of Fig.~\ref{fig:THDM_HHxs_differential} we compare the differential cross-section computed using tree-level (dotted black) and loop-corrected trilinear couplings with (solid red) and without (dashed blue) momentum effects taken into account. The curve with tree-level trilinear couplings is identical to the SM prediction, while both curves including loop corrections to the trilinear couplings show a drastic difference in the shape. On the one hand, we observe a large enhancement at the di-Higgs threshold as well as a dip that occurs near the di-top threshold, which is due to the cancellation of the triangle and box form factors that in the SM happens exactly at the kinematic threshold, and originates because of the shift of $\kappa_{\lambda}$ from 1 at tree-level to 1.7 at the one-loop level. On the other hand, a peak appears at the resonant mass $m_{hh} \sim m_{H} \sim 600$ GeV due to the change in $\lambda_{hhH}$ from zero at tree-level to $\sim335$ GeV at one loop. Once the loop corrections to the trilinear couplings are included, the total cross-section (shown in the legend) is enhanced by almost a factor of two. The inclusion of momentum effects in the result with one-loop trilinear couplings does not yield a significant modification in the shape of the distribution. However, the total cross-section is reduced by $\sim$25\% compared to the effective coupling approximation. Thus the bulk of the corrections is captured by the zero-momentum one-loop effective trilinear coupling. This result highlights the importance of including at least one-loop corrections to the trilinear Higgs couplings when considering scalar extensions of the SM with heavy new physics~\cite{Heinemeyer:2024hxa} (see also Ref.~\cite{Egle:2026def} for recent work on the inclusion of the dominant two-loop corrections to trilinear scalar couplings in this context).

In the right panel of Fig.~\ref{fig:THDM_HHxs_differential} we show the same quantities as in the left plot but for a different scenario. In this case all the physical scalar masses and the BSM mass parameter $M$ are set to 400 GeV. Since all masses are degenerate, the corrections to the trilinear scalar couplings arise only from SM particles and not from the BSM scalars. This is a consequence of the large loop corrections being governed by couplings proportional to the BSM mass splittings (\textit{c.f.} Refs.~\cite{Altmann:2025feg,Bahlproc:2026abc}
). Therefore, the impact of loop corrections to the trilinear couplings is much smaller in this case: only a small deviation from the SM is visible at an invariant mass of $m_{hh} \sim $ 400 GeV, seen as a small dip in the cross-section. For the case where the momentum dependence is neglected  
a dip-peak structure arises. It should be noted, however, that such effects will be largely smeared out by the limited experimental resolution and binning, while in the displayed plot no experimental smearing is applied~\cite{Arco:2022lai,Heinemeyer:2024hxa,Arco:2025nii}.
The total cross-section in this scenario
decreases by slightly less than 10\% as a consequence of the momentum effects, which is smaller than other sources of theory uncertainty in the Higgs pair production process. 

Additionally, for the THDM we have explored the impact of the renormalisation scheme choice, making use of the semi-automatic renormalisation procedure implemented in \texttt{anyBSM}. Specifically, we have modified the KOSY condition on the BSM mass parameter $M$, 
namely an $\overline{MS}$ renormalisation, 
to an OS counterterm through OS conditions on one of the trilinear couplings (similar to a  scheme devised for the general singlet extension of the SM in Ref.~\cite{Braathen:2025qxf}). The scheme dependence of the results for the trilinear couplings can be used as an estimate of the size of the unknown higher-order BSM corrections. We find that there is a relatively small uncertainty from unknown higher-order corrections for couplings involving identical Higgs bosons, such as $\lambda_{hhh}$ and $\lambda_{HHH}$. However for $\lambda_{hhH}$ and $\lambda_{hHH}$, there is a significantly larger dependence on the renormalisation scheme in comparison to the overall size of the coupling, indicating a larger uncertainty on the prediction at the one-loop order, in particular for large values of $M$. At the level of the total Higgs pair production cross-section, we find a significant uncertainty for $M>550$ GeV, indicating the need for more precise predictions including higher-order BSM contributions.

As a second example we consider models with multiple scalar resonances, in particular the NTHDM~\cite{Heinemeyer:2021msz}, which extends the type II THDM by a real singlet, the STHDM~\cite{Biekotter:2021ovi,Biekotter:2022bxp}, which extends the type II THDM by a complex singlet and the TRSM~ \cite{Robens:2019kga}, which extends the SM by two-real singlets.

The example shown for the NTHDM is allowed by all theoretical and experimental constraints and corresponds to the parameters  $\alpha_1 = 1.019$, $\alpha_2 = -0.076$, $\alpha_3 = 0.960$, $
M_{h_1}  = \unit[125]{GeV}$, $ M_{h_2} = \unit[695]{GeV}$, $ M_{h_3} = \unit[773]{GeV}$, $\label{eq:N1}
M_{A} = \unit[673]{GeV}$, $  M_{H^\pm} =\unit[762]{GeV}$, $ M_{12}=\unit[367]{GeV}$, $\tan\beta=1.53, $ $ v_S=\unit[415.4]{GeV}$ (see Ref.~\cite{Bahl:2023eau} for details on the model conventions and notations). We generalise this point to the other two models by  setting the relevant parameters to those values. In the following, $h_1$ is identified with the Higgs boson discovered at 125 GeV. In addition in the STHDM we chose $M_{\chi}=\unit[63]{GeV}$ (which does not affect most constraints) and keep the rest of the parameters equal to the N2HDM. In the TRSM, we set the two singlet VEVs equal to the value of the singlet VEV of the benchmark point in the NTHDM, $v_S^{\rm TRSM}=v_X^{\rm TRSM}=v_S^{\rm NTHDM}$. For simplicity, the \ensuremath{\mathbb{Z}_2}\xspace-breaking parameters of the TRSM are set to zero. In this scenario, the BSM sector of the TRSM is fully specified by the masses and VEVs of the singlets and three scalar mixing angles, which we identify with their N2HDM counterparts according to the type of mixing (SM-BSM or BSM-BSM mixing) that they govern.

\begin{figure}
    \centering
    \includegraphics[width=0.49\textwidth]{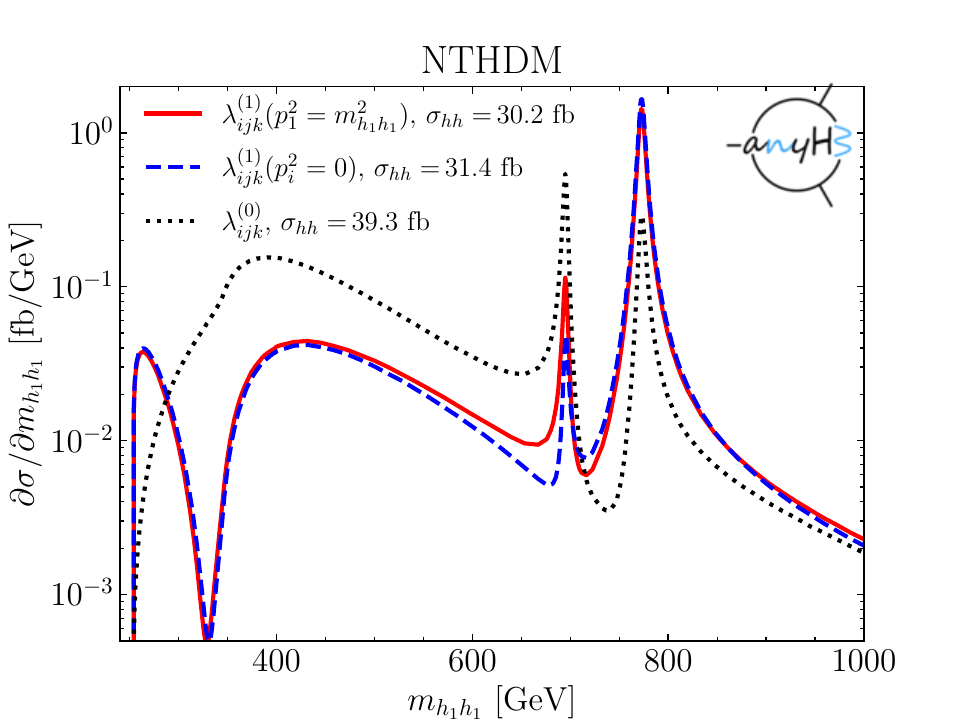}
    \caption{Differential cross-section as a function of the di-Higgs invariant mass for a scenario with two heavy CP-even resonances in the NTHDM. The chosen masses and mixing angles for the considered benchmark point are introduced in the text.}
    \label{fig:doublepeaks_NTHDM}
\end{figure}

\begin{figure}
    \centering
    \includegraphics[width=0.49\textwidth]{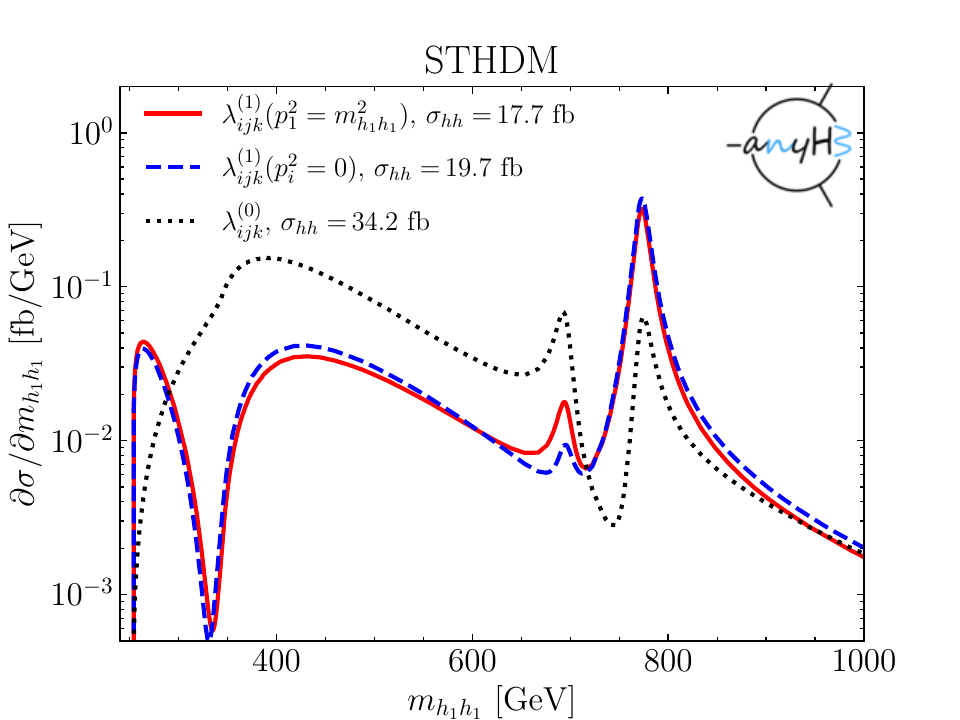}
        \includegraphics[width=0.49\textwidth]{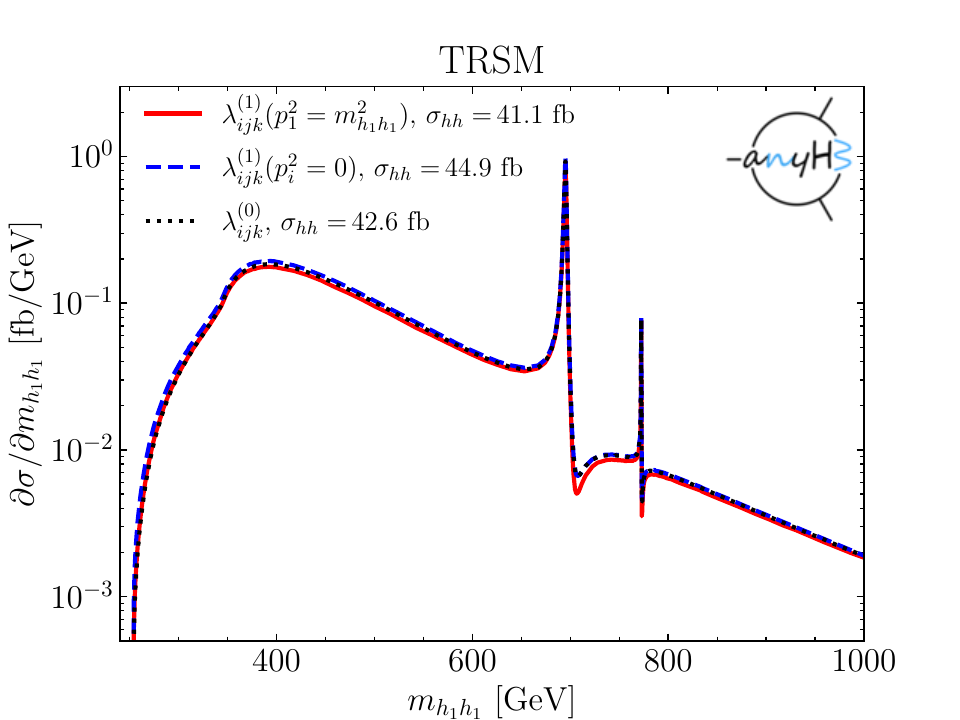}
    \caption{Differential cross-sections as a function of the di-Higgs invariant mass for scenarios with two heavy CP-even resonances in the STHDM (left) and the TRSM (right). The chosen masses and mixing angles, adapted from the benchmark point in the NTHDM, are described in the text.}
    \label{fig:doublepeaks_STHDM_TRSM}
\end{figure}

In all scenarios considered here the top Yukawa couplings of the SM-like Higgs boson are very close to the SM value, while the top-Yukawa couplings of the BSM Higgs bosons are suppressed with respect to the SM-like Higgs Yukawa coupling. However, the BSM decay widths are substantially larger than the SM-like Higgs decay width, mainly due to the decay to top quarks being kinematically allowed and, in case of the STHDM, the additional decay channel into two dark matter candidates. 

Loop corrections enhance the $\lambda_{111}$ and $\lambda_{113}$ 
couplings by roughly a factor of two, whereas $\lambda_{112}$ is suppressed roughly by a factor of two. In the TRSM, loop corrections have only a minor impact, as a consequence of all couplings being fully determined by the mixing angles, masses and VEVs.

The impact of loop corrections to the trilinear Higgs couplings on the invariant di-Higgs mass distribution in the NTHDM is shown in Fig.~\ref{fig:doublepeaks_NTHDM}. Similar to the THDM case,  $\kappa_{\lambda}$ is modified  from a SM-like value at tree-level to $\kappa_{\lambda} \sim 2$ at one-loop. This correction to $\lambda_{111}$ generates an enhancement at the di-Higgs threshold as well as a dip in the spectrum at $m_{h_1h_1}\sim \unit[330]{GeV}$. In addition, the loop-level suppression of $\lambda_{112}$, from 93.5 GeV at tree-level to 40.3 GeV at one-loop, decreases the height of the first resonance peak around $m_{h_1h_1}\sim \unit[700]{GeV}$, while the second peak at $m_{h_1h_1}\sim \unit[800]{GeV}$ is enhanced due to the loop-induced increase of $\lambda_{113}$, from 160.5 GeV at tree-level to 342.8 GeV at one-loop. These features highlight the relevance of including radiative corrections to the involved trilinear scalar couplings when computing predictions for Higgs pair production. We observe that for the scenario considered here momentum-dependent effects only lead to subdominant modifications, as is generally the case for heavy BSM scalar states.

Finally, the corresponding results for the STHDM and the TRSM are shown in Fig.~\ref{fig:doublepeaks_STHDM_TRSM}. The overall behaviour in the STHDM (left panel) closely resembles the one observed in the NTHDM since the two models mainly differ by the presence of the additional dark matter candidate in the STHDM. The corrections to $\kappa_{\lambda}$  are of the same order as in the NTHDM, thus inducing the same dip structure at lower $m_{h_1h_1}$ values. However, the larger decay widths of $h_2$ and $h_3$ in the STHDM compared to the NTHDM lead to noticeably broader resonance peaks; in particular the decay widths have values of 5.1 GeV and 8.2 GeV, respectively, in the N2HDM and of 17.7 GeV and 16.8 GeV in the STHDM. In the TRSM (right panel), the influence of loop-corrected trilinear couplings is, as expected, comparatively small. Since $\lambda_{111}$ is not significantly enhanced ($\kappa_{\lambda} = 0.98$ at tree-level and $\kappa_{\lambda} = 0.93$ at one-loop), the enhancement at the di-Higgs threshold is absent, and no dip appears in the invariant mass distribution. Moreover, the substantially smaller decay widths of $h_2$ and $h_3$ (2.6 GeV and 0.1 GeV, respectively) result in much narrower resonance peaks than in the NTHDM and STHDM cases.

Besides the functionalities of the code explained above, some of the new features include the automatic computation of the decay widths at the leading order using the optical theorem, the calculation of effective coupling modifiers to all scalar couplings in a BSM model with respect to a reference model (the default being the SM) as well as a link of the effective-coupling output of \texttt{anyBSM} to \texttt{HiggsTools}~\cite{Bahl:2022igd}. Further functionalities such as a link to Monte-Carlo generators are currently under development.

\section{Conclusions}

In this work, we have presented major extensions of the \texttt{anyBSM} framework. First, the new version of the \texttt{anyH3} module now allows computations of one-loop corrections to fully general trilinear scalar couplings, including interactions among non-identical scalar mass eigenstates. We have also shown that such corrections can be especially relevant in scalar extensions of the SM. For instance, in scenarios in the alignment limit at tree level, where certain trilinear couplings vanish at tree level, radiative effects can induce resonant contributions in scalar-boson pair production channels which would not be visible in other processes.

Second, we introduced the new subpackage \texttt{anyHH}, which provides predictions for differential and total di-Higgs cross-sections at the (HL-)LHC for the production of two neutral scalar states. The calculation consistently incorporates loop-corrected trilinear couplings and optionally their full momentum dependence. The implementation has been validated against known SM and BSM results in the literature.

We have illustrated a range of phenomenological applications for this new tool. In particular, we have demonstrated the relevance of radiative corrections to trilinear couplings in Higgs pair production in the THDM, leading to the appearance of resonant peaks which would be absent if only tree-level couplings were taken into account. Moreover, we observed that the momentum dependence of the loop-level trilinear couplings can have a significant effect on the total cross-section. We have also shown the versatility of \texttt{anyHH} by generating new results in models with several scalar resonances contributing to the Higgs pair production process. In a broad class of scenarios, loop effects in the trilinear couplings constitute the main source of corrections to the process, and possible effects induced in this way should therefore be taken into account in experimental analyses of di-Higgs production and in direct BSM searches.

Overall, our code provides a versatile and model-independent tool for predicting Higgs pair production in extended Higgs sectors, allowing users to obtain loop-corrected trilinear scalar couplings and automatically and consistently incorporating those in the cross-section prediction for essentially any perturbative BSM scenario described with \texttt{UFO} model files.

\section*{Acknowledgements}
J.B.\ and G.W.\ acknowledge support by the Deutsche Forschungsgemeinschaft (DFG, German Research Foundation) under Germany's Excellence Strategy --- EXC 2121 ``Quantum Universe'' --- 390833306. This work has been partially funded by the Deutsche Forschungsgemeinschaft (DFG, German Research Foundation) --- 491245950. J.B. is supported by the DFG Emmy Noether Grant No.\ BR 6995/1-1. K.R. acknowledges the support of the Spanish Agencia Estatal de Investigación through the grant “IFT Centro de Excelencia Severo Ochoa CEX2020-001007-S”. The project that gave rise to these results received the support of a fellowship from the “la Caixa” Foundation (ID 100010434). The fellowship code is LCF/BQ/PI24/12040018.


\bibliography{bibliography.bib}

\end{document}